\documentclass[english]{article}
\usepackage[T1]{fontenc}
\usepackage[latin9]{inputenc}
\usepackage{amssymb}

\makeatletter

\newcommand{\lyxaddress}[1]{
\par {\raggedright #1
\vspace{1.4em}
\noindent\par}
}

\usepackage{babel}
\makeatother

\begin{document}

\title{Electric Dipole Moment and Neutrino Mixing due to Planck Scale Effects}

\author{Bipin Singh Koranga}

\maketitle

\lyxaddress{Kirori Mal College (University of Delhi,) Delhi-110007, India}

\begin{abstract}
In this paper, we consider the effect of Planck scale operators on
electric dipole moment of the electron $de$. The electric diople
moment of the electron, $de$ is known to vanish up to three loops
in the standard model with massless neutrinos We consider the planck
scale operator on neutrino mixing. We assume that the neutrino masses
and mixing arise through physics at a scale intrimediate between planck
scale and the electroweak braking scale. We also assume, that just
above the electroweak breaking scale neutrino mass are nearly degenerate
and the mixing is bi-maximal. Quantum gravity (Planck scale) effects
lead to an effective $SU(2)_{L}\times U(1)$ invariant dimension-5
Lagrangian symmetry involving Standard Model. On electroweak symmetry
breaking, this operator gives rise to correction to the neutrino masses
and mixings these additional terms can be considered as perturbation
to the bi-bimaximal neutrino mass matrix We assume that the gravational
interaction is flavour blind and we study the neutrino mixing and
electric dipole moment due to the Planck scale effects.
\end{abstract}

\section{Introduction}

THe electric diople moment (e.d.m) of fermions can provide a unique
window to probe into the nature of the force that are responsible
for CP violation {[}1]. experimental limit on the diople moment of
netron have reached the level of $6\times10^{-26}$ ecm {[}2] and
have helped to constrain on theoretical model of CP violation. electric
diploe moment of the electron has severely been constrained by atomic
measurements in Cs $(de\leqslant10^{-26})$and Tl $(de\leqslant10^{-27}ecm)$
{[}3]. The experimental values {[}4,5] are given by

\begin{equation}
dn=(30\pm50)\times10^{-27}e.cm\end{equation}

and

\begin{equation}
de=(1.8\pm1.2\pm1.0)\times10^{-27}e.cm\end{equation}

These values give important constrains on CP violating phase beyond
the Standard Model (S.M). CP violation encoded in the complex element
of the C.K.M Unitary Matrix, V in the quark sector and observable
effects are propertional to the term {[}5]

\[
J=Im\left(V_{ij}V_{kj}^{*}V_{kl}V_{il}^{*}\right)\]

Obiviously J is non vanishing only if not all the elements of $V_{ij}$
can be made real and this implies the existence of at least three
generation of non degenerate massive quarks, standard model $de$
is estimate at the four loop lecvel given by {[}6]

\begin{equation}
de\sim\frac{eG_{F}}{\pi^{2}}\left(\frac{\alpha}{2\pi}\right)^{3}meJ\leq4\times10^{-38},\end{equation}

where $J\leq4\times10^{-38},$ for the C.K.M element is used eq(3),
indicates the opportunity that $de$ is a clean test of CP violation
beyond the S.M. Now the leptonic sector willl exhibit the C.K.M type
of mixing as the analoguge of the C.K.M nmatrix in the quark sector.
In this paper, we assume the matrix are massive Majorana particle.
We consider $de$as induced by a leptonic C.K.M mechanisim which results
from such massive majorana neutrino mixing. The interaction lagragian
involving the electric diploe moment is given by {[}7]

\begin{equation}
L=-\frac{1}{2}deF_{\mu\nu}\overline{e}i\sigma^{\mu\nu}\gamma_{5}e,\end{equation}

and the charge leptonic current interaction is

\begin{equation}
\frac{1}{\sqrt{2}}W^{\mu}U_{ij}\overline{\nu_{i}}\gamma_{\mu}\frac{1-\gamma_{5}}{2}e_{j}+hc,\end{equation}

where $U_{ij}$is the charged current mixing matrix analoge to the
C.K.M matrix in the quark sector. The importance of mixing matrix
U is unitary and the elements are in general complex. In the case
of Majarona neutrino, after some satandrd manimulation. de can be
mwrite in terms of Feynmen intergals is given by {[}7]\begin{equation}
de=\pm\frac{e\alpha^{2}me}{256\pi^{2}s_{w}^{4}}m_{i}m_{j}J_{ij}^{l}F,\end{equation}

Inntergal function F and the CP violating factor $J$ is given by

\begin{equation}
F=\frac{x(1-x)^{2}[(1-s)^{2}-(t+u)^{2}]+xy^{2}u(1-u)-x(1-x)y(1+3s+t+u-2tu-2u^{2})}{m_{i}^{2}x(10x)(1-s-t-u)+m_{j}^{2}(1-x-y)u+m_{w}^{2}(x(1-x)(s+t)+yu+m_{i}^{2}xu)^{2}},\end{equation}

\begin{equation}
J_{ij}^{l}=Im(U_{je}^{*}U_{jl}^{*}U_{il}U_{ie}),\end{equation}

where $l=e,\,\mu,\,\tau$ for the S.M

Goldstone boson exchange digaram are less important. We now give a
semi quantitive estimate of $de$ is given by {[}8]

\begin{equation}
de\sim\frac{\alpha^{2}mem_{i}m_{j}(m_{i}^{2}-m_{j}^{2})}{256\pi^{2}s_{w}^{2}m_{w}^{6}}J_{ij}F(m_{l}^{2}/m_{w}^{2},m_{i}^{2}/m_{w}^{2},m_{j}^{2}/m_{w}^{2})(1.97\times10^{-16})e-cm,\end{equation}

where all masses are taken in unit of Gev. The G.I.M factor $(m_{i}^{2}-m_{j}^{2})/m_{w}^{2}$
for neutrino are explicit the fact that $de$ vanishes, where neutrino
are massless. We can see from eq(3), eq(6), expression of electric
diploe moment (E.D.M) proportional of factor J. Due to Planck scale
effects, E.D.M proportonal to the jJarlskog determiant

\begin{equation}
de^{}\propto J_{ij}^{l},\end{equation}

where $J_{ij}^{'}$is related to mixing angle 

Electric dipole moment and Neutrino Mixing Angles due to Planck Scale
Effects are given in section 2. In section 3 gives the results on
electric dipole moment and neutrino mixing.

\section{Electric Dipole Moment and Neutrino Mixing Angles due to Planck Scale
Effects}

The calculation developed in an earlier paper {[}9]. A natural assumption
is that unperturbed ($0^{th}$ order mass matrix) $M$~is given by

\begin{equation}
\mathbf{M}=U^{*}diag(M_{i})U^{\dagger},\end{equation}

where, $U_{\alpha i}$ is the usual mixing matrix and $M_{i}$ , the
neutrino masses is generated by Grand unified theory. Most of the
parameter related to neutrino oscillation are known, the major expectation
is given by the mixing elements $U_{e3}.$ We adopt the usual parametrization.

\begin{equation}
\frac{|U_{e2}|}{|U_{e1}|}=tan\theta_{12},\end{equation}

\begin{equation}
\frac{|U_{\mu3}|}{|U_{\tau3}|}=tan\theta_{23},\end{equation}

\begin{equation}
|U_{e3}|=sin\theta_{13}.\end{equation}

In term of the above mixing angles, the mixing matrix is

\begin{equation}
U=diag(e^{if1},e^{if2},e^{if3})R(\theta_{23})\Delta R(\theta_{13})\Delta^{*}R(\theta_{12})diag(e^{ia1},e^{ia2},1).\end{equation}

The matrix $\Delta=diag(e^{\frac{1\delta}{2}},1,e^{\frac{-i\delta}{2}}$)
contains the Dirac phase. This leads to CP violation in neutrino oscillation
$a1$ and $a2$ are the so called Majoring phase, which effects the
neutrino less double beta decay. $f1,$ $f2$ and $f3$ are usually
absorbed as a part of the definition of the charge lepton field. Planck
scale effects will add other contribution to the mass matrix that
gives the new mixing matrix can be written as {[}9]

\[
U^{'}=U(1+i\delta\theta),\]

\begin{equation}
=\left(\begin{array}{ccc}
U_{e1} & U_{e2} & U_{e3}\\
U_{\mu1} & U_{\mu2} & U_{\mu3}\\
U_{\tau1} & U_{\tau2} & U_{\tau3}\end{array}\right)+i\left(\begin{array}{ccc}
U_{e2}\delta\theta_{12}^{*}+U_{e3}\delta\theta_{23,}^{*} & U_{e1}\delta\theta_{12}+U_{e3}\delta\theta_{23}^{*}, & U_{e1}\delta\theta_{13}+U_{e3}\delta\theta_{23}^{*}\\
U_{\mu2}\delta\theta_{12}^{*}+U_{\mu3}\delta\theta_{23,}^{*} & U_{\mu1}\delta\theta_{12}+U_{\mu3}\delta\theta_{23}^{*}, & U_{\mu1}\delta\theta_{13}+U_{\mu3}\delta\theta_{23}^{*}\\
U_{\tau2}\delta\theta_{12}^{*}+U_{\tau3}\delta\theta_{23}^{*}, & U_{\tau1}\delta\theta_{12}+U_{\tau3}\delta\theta_{23}^{*}, & U_{\tau1}\delta\theta_{13}+U_{\tau3}\delta\theta_{23}^{*}\end{array}\right).\end{equation}

Where $\delta\theta$ is a hermition matrix that is first order in
$\mu${[}9,10]. The first order mass square difference $\Delta M_{ij}^{2}=M_{i}^{2}-M_{j}^{2},$get
modified {[}9,10] as

\begin{equation}
\Delta M_{ij}^{'^{2}}=\Delta M_{ij}^{2}+2(M_{i}Re(m_{ii})-M_{j}Re(m_{jj}),\end{equation}

where

\[
m=\mu U^{t}\lambda U,\]

\[
\mu=\frac{v^{2}}{M_{pl}}=2.5\times10^{-6}eV.\]

The change in the elements of the mixing matrix, which we parametrized
by $\delta\theta${[}19], is given by

\begin{equation}
\delta\theta_{ij}=\frac{iRe(m_{jj})(M_{i}+M_{j})-Im(m_{jj})(M_{i}-M_{j})}{\Delta M_{ij}^{'^{2}}}.\end{equation}

The above equation determine only the off diagonal elements of matrix
$\delta\theta_{ij}$. The diagonal element of $\delta\theta_{ij}$
can be set to zero by phase invariance. Using Eq(8), we can calculate
neutrino mixing angle due to Planck scale effects,

\begin{equation}
\frac{|U_{e2}^{'}|}{|U_{e1}^{'}|}=tan\theta_{12}^{'},\end{equation}

\begin{equation}
\frac{|U_{\mu3}^{'}|}{|U_{\tau3}^{'}|}=tan\theta_{23}^{'},\end{equation}

\begin{equation}
|U_{e3}^{'}|=sin\theta._{13}^{'}\end{equation}

For degenerate neutrinos, $M_{3}-M_{1}\cong M_{3}-M_{2}\gg M_{2}-M_{1},$
because $\Delta_{31}\cong\Delta_{32}\gg\Delta_{21}.$ Thus, from the
above set of equations, we see that $U_{e1}^{'}$ and $U_{e2}^{'}$
are much larger than $U_{e3}^{'},\,\, U_{\mu3}^{'}$ and $U_{\tau3}^{'}$.
Hence we can expect much larger change in $\theta_{12}$ compared
to $\theta_{13}$ and $\theta_{23}.$ As one can see from the above
expression of mixing angle due to Planck scale effects, depends on
new contribution of mixing $U^{'}=U(1+i\delta\theta).$ Due to Planck
Scale Effects, new proportional factor of electric dipole moment (E.D.M)
is given by

\begin{equation}
de^{'}\propto J_{ij}^{'l}\propto Im(U_{je}^{'*}U_{jl}^{'*}U_{il}^{'}U_{le}^{'}),\end{equation}

where $U_{je}^{'},$$U_{jl}^{'},$$U_{il}^{'}$ and $U_{ie}^{'}$
is the mixing angle parameter given by new mixing ( $U^{'}=U(1+i\delta\theta)$)due
to Planck scale Effects

\section{Results and Discussions}

We assume that, just above the electroweak breaking scale, the neutrino
masses are nearly degenerate and the mixing are bimaximal, with the
value of the mixing angle as $\theta_{12}=45^{o},\,\theta_{23}=45^{o}$
and $\theta_{13}=0.$ Taking the common degenerate neutrino mass to
be 2 eV, which is the upper limit coming from tritium beta decay {[}11].
We compute the modified mixing angles using Eqs (11)-(13). We have
taken $\Delta_{31}=0.002eV^{2}[12]$ and $\Delta_{21}=0.00008eV^{2}${[}13].
For simplicity we have set the charge lepton phases $f_{1}=f_{2}=f_{3}=0.$~Since
we have set the $\theta_{13}=0,$ the Dirac phase $\delta$~drops
out of the zeroth order mixing angle. We consider the Planck scale
effects on neutrino mixing and we get the given range of mixing parameter
of MNS marix

\begin{equation}
U^{'}=R(\theta_{23}+\epsilon_{3})U_{phase}(\delta)R(\theta_{13}+\epsilon_{2})R(\theta_{12}+\epsilon_{1})\end{equation}

In Planck scale, only $\theta_{12}$have resonable deviation {[}13]
and $\theta_{13},$$\theta_{23}$deviation is very small less than
$0.3^{o}${[}12]. In the new mixing due to Planck scale effects, we
get the new multiplicity factor of electric dipole moment , which
is proprtonal to Jarlskog determiant and Planck scale mixing chang
the Jarlskog Determiant{[}14]. If there exist Majorana neutrino with
masses $m_{i}^{2}\leq m_{w}^{2}$ and $|U_{ei}|^{2}\leq10^{-2}$,obtained
from the charged current universality constraints {[}15]. We get $de\leq10^{-32}e-cm$assuming
that the intergal function F in eq(9) is of the order unity.

\section{Conclusions}

In conclusions, we find that in genral majorana neutrino can induce
electric dipole moment at the two loop level. Numerical estimate for
$de\leq10^{-32}e-cm$ even for Majorana neutrino of masses in the
100 GeV range. The electric diople moment factor in eq(6) can be effected
by mixing parameter $\theta_{12,}$ $\theta_{23}$ and $\theta_{13}$.
We consider the main part of neutrino mixing arise from GUT scale
operator. We further assume that GUT scale symmetery the neutrino
mixing to be bi-maximal. We compute the first first order correction
to neutrino mass eigen value and mixing angles. In {[}10], it was
shown that the change in $\theta_{13\,}due\,$to perturbation is small.
We also show that the change in $\theta_{23}$ also is small (less
then $3^{o}$) only change in $\theta_{12}$can be substantial ( about
$\pm3^{o}).$

In this paper, we study electric diploe moment due to Planck scale
effects. For majorana neutrino with three flavour, the expression
is $de\sim J_{ij}^{l}F.$In this paper, finally we wish make a important
comment due to Planck scale effects mixing angle $\theta_{12}$change
effectively the electric dipole moment (E.D.M) and magnetic moment
{[}15]. Non zero value of electric dipole moment in eq(6) also indicate
the existence of E.D.M, which is related to CP violation..

\end{document}